# Agglomeration of Dust Particles in the Lab

Lorin S. Matthews, Jorge Carmona-Reyes, Victor Land, and Truell W. Hyde

*Center for Astrophysics, Space Physics, and Engineering Research*
*Baylor University, One Bear Place 97310, Waco, Texas 76798-7310 USA*

**Abstract.** Dust aggregates are formed in a laboratory plasma as monodisperse spheres are accelerated in a self-excited dust density wave. The asymmetric charge on the aggregates causes them to rotate as they interact with the sheath electric field or other aggregates The charge and dipole moment can be estimated and compared to numerical models. "Dust molecules", where two particles are electrostatically bound but not physically touching, are also observed.

**Keywords:** dusty plasma, coagulation, dust charging

**PACS:** 52.27.Lw, 94.05.Bf, 95.30.Wi, 96.30.Vb

## METHOD

Aggregates were formed in the lab following the method of Du *et al* [6]. The experiments were performed in a modified GEC rf reference cell, which has a lower electrode driven at 13.56 MHz and a grounded upper electrode, separated by 2.54 cm. A glass box was placed on the lower electrode to provide additional horizontal confinement. Single 8.93 µm gold-coated melamine formaldehyde micro-particles were injected into an argon discharge from a shaker at the top of the cell. With the plasma maintained with 80 V peak-to-peak voltage (measured just before the shunt circuit) and pressure of 500 mTorr, the particles formed a stable cloud within the box. A butterfly valve was used to rapidly decrease the pressure in the cell to 50 mTorr, which induced dust density waves, increasing the relative velocities between particles and inducing agglomeration. Upon increasing the pressure to 500 mTorr, aggregates could be imaged below the main dust cloud. The dust cloud and coagulates were back-illuminated with a 500 W flood lamp and the aggregates imaged using a CMOS monochromatic high speed camera (1024FASTCAM Photron with microscope lens attached) at 3000 fps.

## RESULTS

Aggregates could be seen to rotate due to the torque caused by the electric field within the sheath (Fig. 1a) or by interacting with other charged aggregates (Fig. 2). Taking advantage of the rotation of the aggregates, a three-dimensional model of the aggregate could be reconstructed (Fig. 1b), allowing physical parameters such as the mass and inertia to be calculated. These parameters, along with the time-resolved dynamic response of the aggregate, allow the charge and dipole moment of the aggregate to be extracted for comparison with numerical models [2].

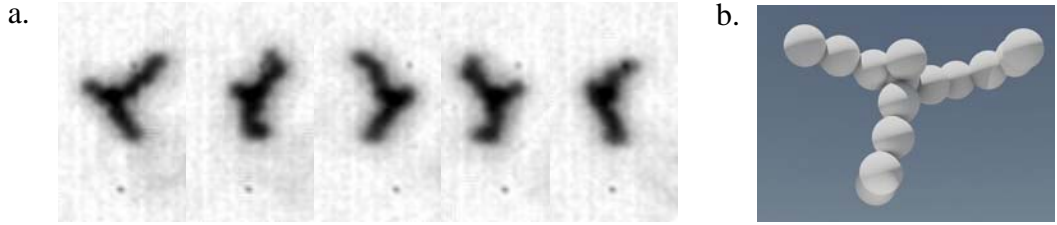

**FIGURE 1.** a) Sequence of images showing a single cluster consisting of fourteen 8.93 μm gold-coated mf particles. b) Reconstructed 3D model of aggregate.

A new interesting phenomena, dust molecules, was also observed (Figures 2 and 3). Two distinct particles could be seen to be in very close proximity, clearly separated, but maintaining almost fixed positions with respect to each other. The apparent bonding was quite robust, with the aggregate molecules remaining bound even when interacting with a second aggregate. The secondary bound aggregates are near the top of the main aggregate, which implies that the bonding may be due to the combination of the ion drag force with a local potential minima of the aggregate structure.

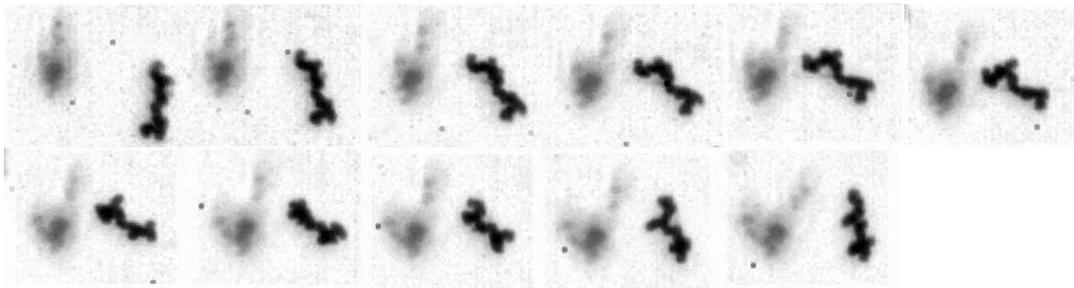

**FIGURE 2.** A sequence of images showing the interaction of two dust aggregates ($\Delta t = 1.67 \times 10^{-3}$s). The aggregate on the right approaches the second aggregate and changes its direction of rotation. Note that the aggregate on the left is actually a bound "dust molecule" with the upper half not physically attached to the lower half. The two pieces remain bound even through the interaction.

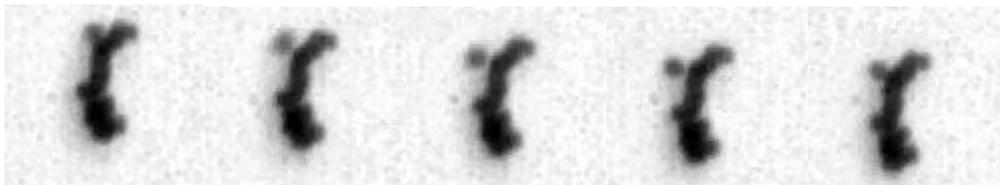

**FIGURE 3.** Sequence of consecutive images showing a single monomer at the upper left which is not physically attached to the aggregate but appears to be electrostatically bound ($\Delta t = 3.33 \times 10^{-4}$s).

## ACKNOWLEDGMENTS

This work is supported by the National Science Foundation under Grant No. 0847127.